%% file: NuMuBarCC_Inclusive_Paper_2024.tex
\documentclass[aps,prl,reprint, preprintnumbers, superscriptaddress]{revtex4-1}

\usepackage{graphicx}   
\usepackage{dcolumn}    
\usepackage{bm}         
\usepackage{lineno}     
\usepackage{amsmath}
\usepackage{xspace}
\usepackage{sidecap}
\usepackage{color}      
\usepackage{comment}    
\usepackage{multirow}
\usepackage{hyperref}   

\newcommand{\dcp}{\ensuremath{\delta_{\text{CP}}}\xspace}
\newcommand{\tmu}{\ensuremath{T_{\mu}}\xspace}
\newcommand{\cosmu}{\ensuremath{\cos{\theta}_{\mu}}\xspace}
\newcommand{\Eavail}{\ensuremath{E_{\text{avail}}}\xspace}
\newcommand{\qsqr}{\ensuremath{Q^{2}}\xspace}
\newcommand{\enu}{\ensuremath{E_{\nu}}\xspace}
\newcommand{\numubarcc}{\ensuremath{\bar{\nu}_{\mu}\ \text{CC}}\ \xspace}
\newcommand{\numucc}{\ensuremath{\nu_{\mu}\ \text{CC}}\ \xspace}
\newcommand{\numubar}{\ensuremath{\bar{\nu}_{\mu}}\xspace}
\newcommand{\numu}{\ensuremath{\nu_{\mu}}\xspace}
\newcommand{\nuebarcc}{\ensuremath{\bar{\nu}_{e}\ \text{CC}}\ \xspace}
\newcommand{\nuecc}{\ensuremath{\nu_{e}\ \text{CC}}\ \xspace}

\begin{document}

\preprint{FERMILAB-PUB-26-0080-PPD}
\title{First inclusive triple-differential measurement of the muon-antineutrino charged-current cross section using the NOvA Near Detector}

\input{novatriplecross2024}


\begin{abstract}
We present the first measurement of the triple-differential muon antineutrino charged-current inclusive cross section, using the NOvA Near Detector and $12.5 \times 10^{20}$ protons on target in the NuMI beam. This sample of muon antineutrino interactions is the largest ever published, with approximately 1 million selected muon antineutrino events.
The triple-differential cross section is measured in the final-state kinetic energy, the scattering angle, and the available energy of the interaction. The measurement enables phase-space regions populated by differing neutrino reaction processes to be isolated and the transition regions between them to be defined. 
The results are compared with the predictions of the main event generators used in the neutrino community and we observe energy- and angle-dependent discrepancies across a broad range of energies and interaction types.
\end{abstract}

\maketitle


\clearpage

Neutrinos are known to change flavor as they propagate from their source to their detection point, a phenomenon known as neutrino oscillations. Long-baseline neutrino experiments~\cite{ref:PhysRevD.106.032004, ref:Eur.Phys.J.C83.782, ref:PhysRevLett.115.111802, ref:PhysRevD.97.072001, ref:PhysRevLett.89.011301,ref:DUNETDR, ref:HyperK} compare energy spectra near the source and after propagation to infer neutrino oscillation probabilities, thereby constraining oscillation parameters, including the mixing angles, the neutrino mass splittings, and the CP-violating phase, \dcp. Comparisons between neutrino- and antineutrino-mode measurements provide enhanced sensitivity to the oscillation parameters. Interpreting the oscillation probabilities requires a thorough understanding of the initial (anti)neutrino flux, detector efficiencies, and the modeling of neutrino--nucleus cross sections. 

Among the leading sources of systematic uncertainty in long-baseline neutrino oscillation measurements are those stemming from neutrino--nucleus interaction modeling~\cite{ref:PhysRevD.106.032004,ref:Eur.Phys.J.C83.782}. Next-generation long-baseline experiments~\cite{ref:DUNETDR, ref:HyperK} will use GeV-scale beam energies and their measurements will be more sensitive to neutrino-nucleus scattering uncertainties than any previous neutrino experiment. 
Moreover, the modeling of nuclear effects in (anti)neutrino--nucleus scattering is a significant challenge in this energy range. Antineutrino--nucleus cross-section measurements, such as the one reported here, constrain models of antineutrino interactions and the complex nuclear environment in which these interactions occur.

Inclusive neutrino cross-section measurements are a comprehensive check of our interaction and nuclear models, as they test the sum of the effects of interaction processes: quasi-elastic (QE), resonant (Res), shallow- and deep-inelastic scattering (S/DIS), and such nuclear effects as two-particle-two-hole (2p2h) and final-state interactions (FSI)~\cite{ref:NuSTECwhitePaper}. NOvA probes a region containing all these interactions, hence an inclusive cross-section measurement by NOvA is sensitive to all of these processes. 

While a variety of \numu charged-current (CC) inclusive double and triple differential measurements are available, muon antineutrino (\numubar) CC inclusive differential measurements have mainly been performed as a function of antineutrino energy or single differential in the antimuon kinematics~\cite{ref:ArgoNeut_CCinclusive,ref:MINOS_CCinclusiveRatio,ref:MINERVA_antineutrinoratio_LE,ref:MINERVA_antineutrinoratio_ME,ref:T2K_CCinclusiveRatio,ref:NINJA_CCinclusiveRatio}. NOvA has previously released a systematics-limited double differential measurement of \numucc inclusive cross sections in the final-state muon kinematics with more than 170 bins~\cite{ref:PhysRevD.107.052011}. This letter augments those results with the first triple differential measurement of \numubarcc inclusive cross sections as a function of final-state antimuon kinematics and available energy (\Eavail, a proxy for visible hadronic energy, defined later in the paper) with 459 bins and approximately 1M selected antineutrino events. This is the highest-statistics measurement to date, enabling model comparisons of the combined leptonic and hadronic final state systems in the 1--5\,GeV energy range across a wide range of reaction processes.

NOvA is a long-baseline neutrino experiment with two detectors designed to measure neutrino oscillations~\cite{ref:PhysRevD.106.032004}. The NOvA detectors are exposed to the NuMI~\cite{ref:numi_fnal} (anti)neutrino beam at Fermilab. The NOvA Near Detector (ND) is located 100\,m underground and 1\,km from the neutrino production target at Fermilab. The ND is placed off-axis from the center of the NuMI beam such that the (anti)neutrino beam energy peaks at around 2\,GeV. The detector is constructed of  alternating horizontal- and vertical-oriented planes of extruded PVC cells. Each cell has a length of 6.6\,cm in the direction of the beam, 3.9\,cm transverse to the beam, and is 3.9\,m long. The cells are filled with a mix of $\sim$95\% mineral oil and $\sim$5\% pseudocumene as scintillator \cite{NOvA_scintillator}. 
A muon catcher, located downstream of the fully active part of the ND, is constructed from pairs of planes of cells separated by slabs of steel, designed to stop and measure muons with energies up to 2.5\,GeV. 
The ND fiducial volume is 300 tons, with a composition of 67\% carbon, 16\% chlorine, 11\% hydrogen, 3\% oxygen, and 3\% titanium by mass. 
A charged particle traversing a cell deposits energy, producing scintillation and Cherenkov light, which is captured by a wavelength-shifting optical fiber threaded through the cell and conveyed to an avalanche photodiode (APD) at one end. The output of the APD is digitized using custom-made front-end electronics. Signals above a threshold are recorded in a data buffer. 

Due to its proximity to the beam source,  the NOvA ND receives an intense
flux of (anti)neutrinos. This letter presents data collected in the ND from June 2016 to July 2019, equivalent to $12.5 \times 10^{20}$ protons-on-target (POT) in the NuMI beam.

\textsc{Geant4}  v9.2.p03~\cite{ref:geant4_v410_a} is used to predict the flux of the NuMI beam. We used external hadron production measurements~\cite{ppfx:ext1, Alt:2006fr, Abgrall:2011ae, Barton:1982dg, Seun:2007zz, Tinti:2010zz, Lebedev:2007zz, Baatar:2012fua, Skubic:1978fi, Denisov:1973zv, Carroll:1978hc, Abe:2012av, Gaisser:1975et, Cronin:1957zz, Allaby:1969de, Longo:1962zz, Bobchenko:1979hp, Fedorov:1977an, Abrams:1969jm} within the Package to Predict the Flux (PPFX)~\citep{ref:ppfx_PhysRevD.94.092005} to constrain hadron production due to the interaction of the beam with the target. Neutrino interactions in the detector material are simulated with GENIE 3.0.6~\cite{PhysRevD.104.072009, Andreopoulos_2010, Andreopoulos:2015wxa}. Our chosen GENIE model configuration (\texttt{G18\_10j\_00\_000}) uses the local Fermi Gas model~\cite{bib:Valencia1} to simulate the initial state of the nucleons inside the nuclei of the detector material. QE and 2p2h interactions of neutrinos with the detector material are simulated using the Val\`{e}ncia model \cite{bib:Valencia3}, Res and coherent (Coh) interactions using the Berger--Sehgal model \cite{PhysRevD.76.113004}, and S/DIS interactions using the Bodek--Yang model \cite{BODEK200270}. The final-state interactions of particles traversing the nucleus are simulated using the GENIE hN2018 model~\cite{Andreopoulos:2015wxa}.

The output of GENIE is modified to create a custom NOvA tune by adjusting the 2p2h and FSI models to establish a physically reasonable central value on which systematic uncertainties can be applied.
The 2p2h model is reweighted to better represent the ND data. The weights are parametrized as two bidimensional Gaussian distributions plus a constant in the space of $(q_0, |q|)$, the energy component and the magnitude of the spatial components of the 4-momentum transfer. 
Only the neutrino-mode data are used to fit the parameters, while the same weights are applied to both neutrinos and antineutrinos. The FSI model is adjusted to better agree with pion scattering data~\cite{Allardyce:1973ce, Saunders:1996ic, Meirav:1988pn, Levenson:1983xu, Ashery:1981tq, Ashery:1984ne, PinzonGuerra:2016uae}. The central value is tuned by changing the Mean Free Path value. Then a scale factor is applied to the charge-exchange, QE, and Absorption channels. The above procedure constitutes the ``NOvA Tune v2'' used in this analysis.

\textsc{Geant4} v4.10 \cite{ref:geant4_v410_a} is used to simulate the deposition of energy in the detector by the traversing particles. The production of scintillation and Cherenkov light in the detector cells and its transfer to the electronic readout are simulated using custom software~\cite{Aurisano_2015}.

The signal for this analysis is \numubarcc interactions within the fiducial volume of the detector. We select events with a contained (anti)muon candidate in the final state, using the same techniques developed by the \numu CC inclusive cross-section measurements presented in Ref.~\cite{ref:PhysRevD.107.052011}. These include requiring at least one reconstructed track of good quality, with a vertex in the fiducial volume, and that the event be fully contained with no energy deposited within several centimeters from the detector edges.  As in Ref.~\cite{ref:PhysRevD.107.052011}, the final antimuon candidates are selected using a Boosted Decision Tree (MuonID), trained on energy deposits and the difference in scattering log-likelihood for muon versus pion hypotheses. 
The MuonID output provides excellent separation of the signal from neutral current (NC), \nuecc, and \nuebarcc backgrounds. In the end, we select 1,042,460 data events with an average 90.6\% purity and 32.8\% efficiency estimated via our simulation. The main background comes from \numucc interactions, accounting for 8.6\% of the final selected sample, which is estimated by simulation. The signal sample is estimated to be approximately 32\% QE, 25\% 2p2h, 31\% Res, 8\% S/DIS, and 2\% Coh interactions.

\begin{figure}
    \centering
    \includegraphics[width=0.95\linewidth]{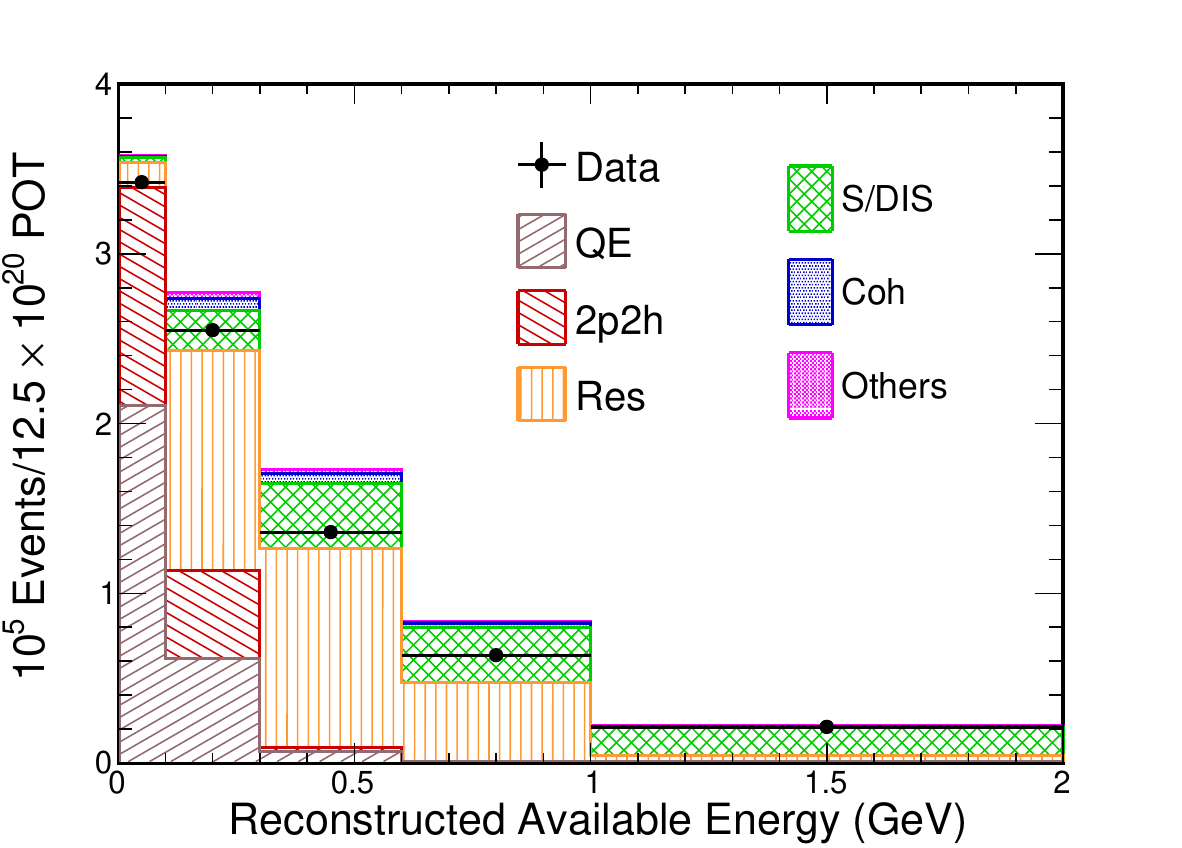}
    \caption{Event distribution of the available energy in the selected sample, broken down by interaction types. Only statistical uncertainties are included in the error bars. The simulation histogram is obtained with GENIE 3.0.6 NOvA Tune v2. } 
    \label{fig:Eavail_interaction_type}
\end{figure}

Our measurement is the triple-differential cross section as a function of the final-state antimuon kinetic energy (\tmu), cosine of final state antimuon angle with respect to the beam direction (\cosmu), and the available energy of interaction, \Eavail. \Eavail is the sum of the total energies of neutral pions and kinetic energies of protons and charged pions in the final state~\cite{ref:Minerva_Eavail_PhysRevLett.116.071802}. It is estimated using a quadratic fit to the reconstructed visible hadronic energy based on simulation.  Antineutrino QE and 2p2h interactions mostly produce neutrons in the final state.  Reflecting this, 48\% of true selected signal events have \Eavail$<$0.1\,GeV. Figure~\ref{fig:Eavail_interaction_type} shows the distribution of \Eavail for the selected sample. This variable is used to separate regions of our measurement that are enhanced in QE and 2p2h, Res, and S/DIS interactions.
The \Eavail$<$0.1\,GeV region is dominated by QE and 2p2h interactions. The region between 0.1 and 0.3\,GeV is characterized by the transition from QE/2p2h to Res processes. At low \tmu, Res processes dominate, whereas at high \tmu, QE/2p2h processes remain dominant. The regions above 0.3\,GeV and up to 1\,GeV are dominated by Res interactions, and above 1\,GeV, S/DIS interactions dominate, without a strong dependence on \tmu. 

The binning of analysis variables is optimized to minimize the migration of events when unfolding them from reconstructed to true space. To keep statistical uncertainties below 7\%, we report only analysis bins with $>$ 200 events.

We apply purity, efficiency, and unfolding corrections derived from our simulation to the selected data.  The purity correction ranges between 60\% and 90\%; bins with lower \Eavail and higher \tmu tend to have higher purity.  The efficiency correction is dominated by containment criteria and ranges between 20\% and 90\%; bins with lower \Eavail, lower \tmu, and forward angles tend to have higher efficiency.
 
We use the iterative D'Agostini method \cite{ref:dagostini} with two iterations in our unfolding procedure. The number of iterations was optimized by minimizing the average Mean Squared Error across all sources of systematic uncertainty. We confirmed that no significant changes to the result were observed after two iterations. 

We assess systematic uncertainties of the measured cross sections by comparing the measurement using our nominal simulation to measurements using systematically shifted simulations for each source of systematic uncertainty, similar to the treatment described in Ref. \cite{ref:PhysRevD.107.052011}. The systemic uncertainties are grouped into categories. Detector response uncertainties include those of detector calibration, production and transport of scintillation and Cherenkov light, and modeling of the energy loss of muons in the detector material. Flux uncertainties comprise those due to hadron production and beamline optics. The dominant uncertainties in hadron production are evaluated by PPFX, which uses a large set of hadron production measurements to better understand and tune the (anti)neutrino flux predictions~\cite{ref:ppfx_PhysRevD.94.092005}. 
Systematic uncertainties resulting from (anti)neutrino--nucleus interaction modeling are determined by modifying various physics parameters in GENIE plus a set of shifts from physics considerations and external data~\cite{ref:NOvA3Flavor}. 
Our neutron systematic uncertainty is derived using an alternative neutron propagation model, \textsc{Menate}~\cite{ref:menate, KOHLEY201259}. Statistical uncertainties are evaluated using Poisson-fluctuated universes and are included in both the data and simulated samples~\cite{ref:PhysRevD.107.052011}. 

Table~\ref{table:systs_uncert} shows the average fractional uncertainties and their bin-to-bin correlations. 
The average fractional systematic uncertainty is dominated by uncertainties in (anti)neutrino flux modeling, which is primarily a normalization effect and has little impact on the shape of the cross section as a function of the observables.  The neutron modeling, detector response, and $\nu$--A modeling uncertainties are typically subdominant, 5\% or less, for \tmu below 1.4\,GeV and \Eavail below 0.6\,GeV. Detector response uncertainties grow to 10--20\% at higher \tmu  and higher \Eavail.  The modeling uncertainties of $\nu$--A are above 10\% in only a few bins at the highest \cosmu. 

\begin{table}[!]
\centering
\caption{Average fractional uncertainties on the triple differential cross section from various sources. Shape-only values are shown in parentheses. 
Uncertainties and correlations are computed as weighted averages, with weights given by the cross section in each bin. 
}
\label{table:systs_uncert}
\resizebox{\columnwidth}{!}{%
\setlength\extrarowheight{2pt}
\begin{tabular}{lccc}
\hline
Source             & Fractional uncertainty (\%)     & Bin-to-bin correlation \\ \hline
Flux               & 9.9 (1.0)                    & 0.99 (0.03)    \\
Detector response  & 6.1 (5.4)                    & 0.21 (0.02)    \\
$\nu$--A            & 2.5 (2.3)                    & 0.17 (0.03)    \\
Neutron            & 2.3 (2.3)                    & 0.04 (-0.01)   \\
Statistical        & 1.4 (1.4)                    & 0.01 (-0.01)   \\
Total              & 12.6 (6.8)                   & 0.69 (0.00)    \\ \hline
\end{tabular}%
}
\end{table}

\begin{figure*}
    \centering
    \includegraphics[width=1.0\linewidth]{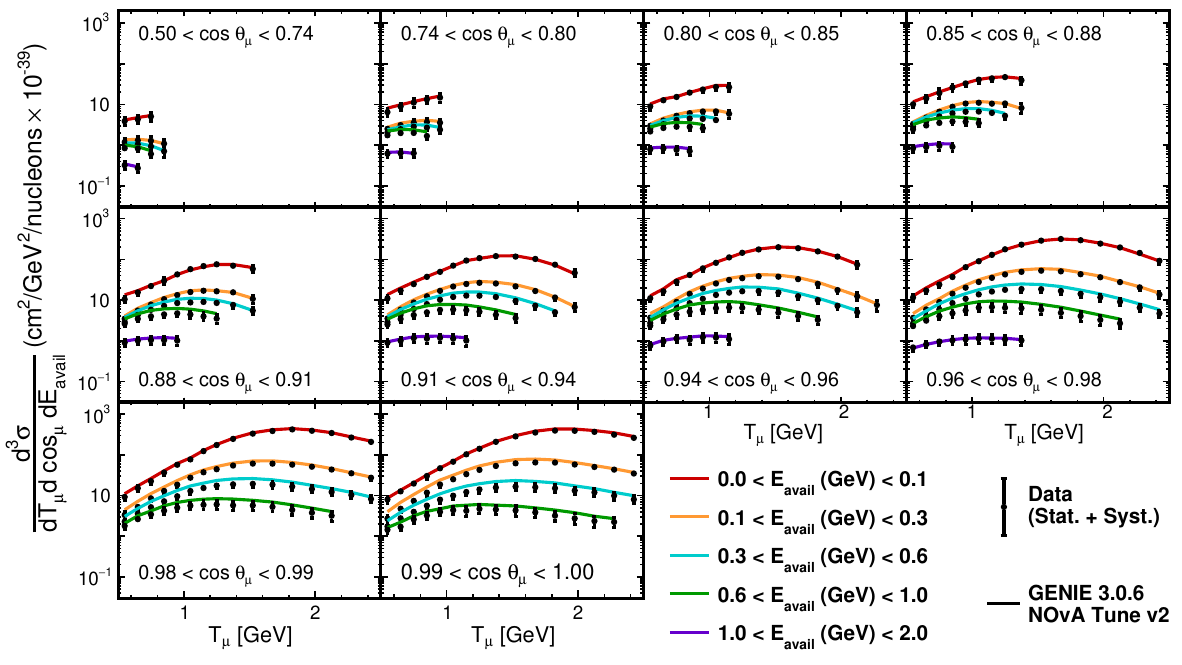}
    \caption{Flux-integrated triple differential cross-section measurements in panels of \cosmu. Error bars include statistical and systematic uncertainties. The predictions from GENIE 3.0.6 NOvA Tune v2  in each \Eavail region are shown in various colors.}
    \label{fig:results_3d}
\end{figure*}

The measured flux-integrated triple differential cross sections are shown in Fig.~\ref{fig:results_3d} vs. \tmu in panels of $\cos\theta_\mu$. The predicted NOvA Tune v2 cross sections are shown by colored curves for the \Eavail regions. 
Despite the NOvA Tune v2 being calculated using neutrino-only data, it models our muon antineutrino measurement well up to the 0.3\,GeV available energy region (red and orange curves), where QE/2p2h interactions dominate. 
Between 0.3 and 1\,GeV (cyan and green curves), our sample is dominated by resonance interactions; here, our measurements are consistently lower than the prediction. We see overall good agreement in the highest \Eavail bin (purple line), where S/DIS dominates. 

\begin{figure}[!h]
    \centering
    \includegraphics[width=0.99\linewidth]{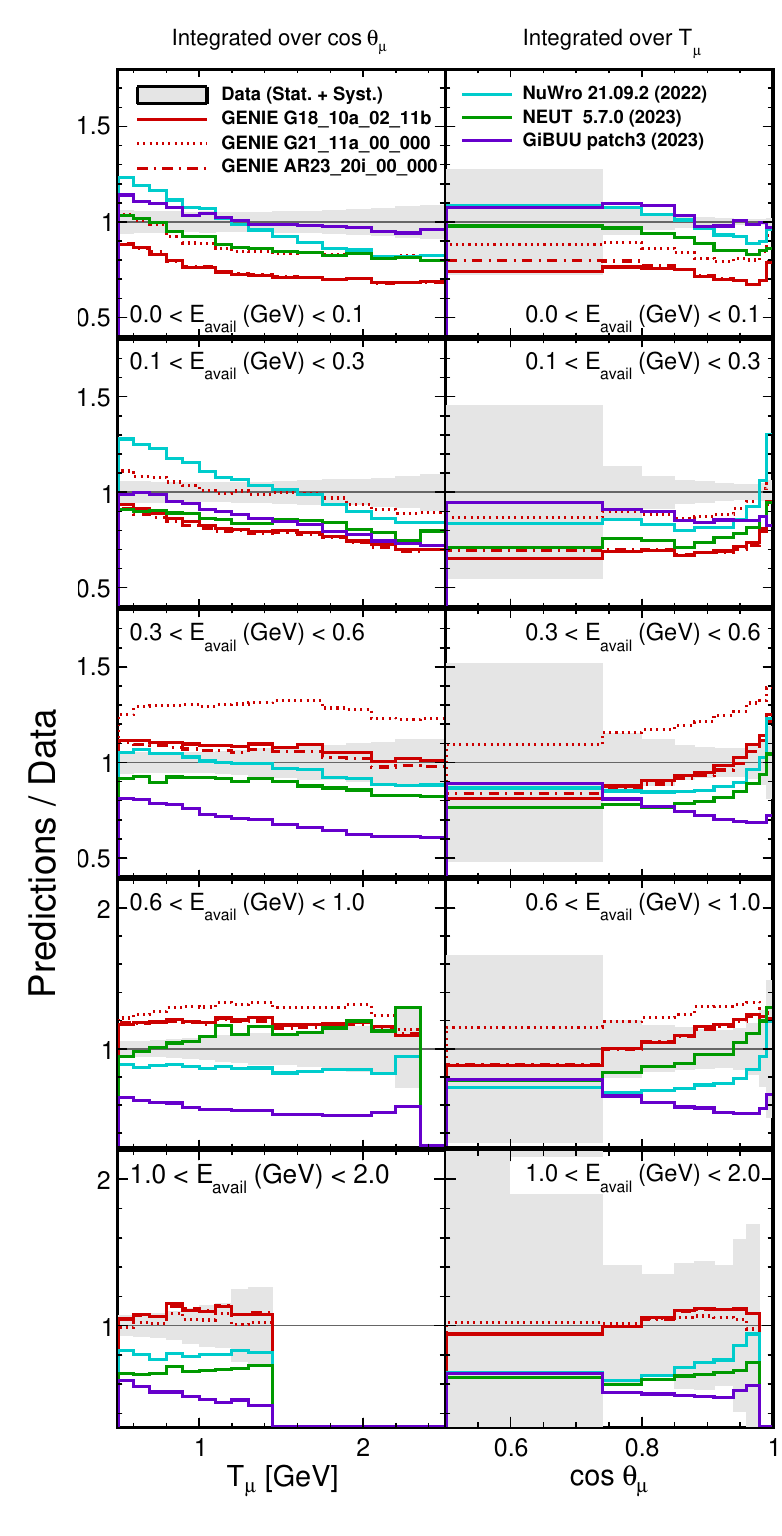}
    \caption{Comparisons of our measurements to various GENIE CMCs, and to the neutrino generators NuWro 21.09.02 (2022) \cite{PhysRevC.86.015505}, NEUT 5.7.0 (2023) \cite{Hayato_2021}, and GiBUU (2023) \cite{BUSS20121, PhysRevD.109.033008}. See Tab.~\ref{tab:generators} for a summary of the models used in these generators. Each plot shows the ratio of the prediction to our measurement. In the left column, the differential cross sections are integrated over \cosmu and shown vs. \tmu. In the right column, the differential cross sections are integrated over \tmu and shown vs. \cosmu. Each row represents one of the \Eavail bins. }
    \label{fig:generator_comparison_integrated}
\end{figure}

Figure~\ref{fig:generator_comparison_integrated} compares our measurements to various GENIE Comprehensive Model Configurations (CMCs), and the neutrino generators NuWro 21.09.02 (2022) \cite{PhysRevC.86.015505}, NEUT 5.7.0 (2023) \cite{Hayato_2021}, and GiBUU (2023) \cite{BUSS20121,PhysRevD.109.033008}. For the GENIE predictions, we consider three separate CMCs:
\texttt{G18\_10a\_02\_11b}, \texttt{G21\_11a\_00\_000}, and \texttt{AR23\_20i\_00\_000}, described in detail in the supplemental material. 
These CMCs were chosen to assess a variety of physics models. 
The left column shows the cross section vs. \tmu, calculated by integrating the three-dimensional cross sections over \cosmu and taking the ratio of the prediction to our measurement. The right column shows the cross section vs. \cosmu, calculated by integrating the cross sections over \tmu and taking the ratio of the prediction to our measurement.  
See the supplemental material for a full comparison of the three-dimensional cross sections to the generator predictions. 

In the very low \Eavail region, below 0.1\,GeV (60\% QE and 35\% 2p2h), all untuned simulations struggle to model our data both in shape and normalization. 
The GENIE Ar23 and G18\_10a models show very similar trends; they have the same QE model (Val\`{e}ncia) but use different 2p2h models (SuSA-v2~\cite{bib:SuSA,bib:SuSAv2} and Val\`{e}ncia~\cite{bib:Valencia1,bib:Valencia3}, respectively), suggesting that the effects of the two 2p2h models are very similar, or that the impacts of the 2p2h models in this slice is smaller than changes in the QE model. The GENIE G21 model, which uses SuSAv2 for both QE and 2p2h, is closer to our measurement, implying that the QE component of the SuSAv2 model agrees better with our data than the QE component of the Val\`{e}ncia model.
Looking at the other neutrino generators in this region, we see small shape discrepancies with GiBUU and NEUT, and some large discrepancies with NuWro, which uses the Llewellyn--Smith model for QE interactions~\cite{bib:LlewellynSmith} and does not include the suppression coming from the random-phase approximation at low momentum.

Between 0.1 and 0.3\,GeV in \Eavail is the transition region from QE+2p2h to Res interactions. All GENIE CMCs compared here have the same Res model (Berger--Seghal). In this transition region, where QE and 2p2h interactions dominate ($\tmu > 1.5$\,GeV), we see a similar trend as in the 0--0.1\,GeV bin of \Eavail, with the QE component of the SuSAv2 model showing better agreement with our data than the Val\`{e}ncia model. Additionally, all generators show shape discrepancies in \cosmu, in particular at the most forward-going angles, where most generator predictions show a sharper predicted increase than we observe in our measurement.  This suggests that additional suppression in the models is needed in this region of antimuon phase space. 

In the 0.3--1.0\,GeV \Eavail region, Res interactions dominate. Here we see \tmu shapes in disagreement across all generators. In particular, apart from GiBUU, all generators use the same resonant model (Berger--Sehgal), suggesting that the model is not capturing the antimuon angular and energy dependence correctly.  Discrepancies of GENIE Ar23 with respect to our measurements in this energy region suggest that DUNE~\cite{ref:DUNETDR} may require additional tuning of their generator. This underscores the need for improved models and options for the Res region in the generators. 

S/DIS interactions dominate \Eavail regions above 1\,GeV. Here, GENIE, which uses the Bodek--Yang model and PYTHIA6~\cite{bib:PYTHIA}, agrees well throughout the phase space (irrespective of which FSI model is used). While NuWro and NEUT also use Bodek--Yang, their FSI tunes differ, resulting in disagreement with measurements, especially at forward angles. GiBUU roughly reproduces the shape of our measurement at high angles, but shows some normalization disagreement at more forward angles. 

The supplemental material presents single differential cross sections vs. antineutrino energy (\enu) and squared momentum transfer (\qsqr), as well as full $\chi^2$ comparisons to the generators.
The data relating to the measurement and systematic uncertainties using covariance matrices can be found at \cite{ref:nova_collaboration_2026_18867876}.

In summary, we have reported the first inclusive triple-differential measurement of the muon-antineutrino charged-current cross section using the NOvA ND, extracted using the largest sample of muon antineutrino interactions (approximately 1 million events) ever published. The cross sections are measured in the final-state \tmu, \cosmu, and \Eavail of the interaction, which allows us to distinguish various neutrino interaction processes and the transition regions between them.
The data are compared with the main event generators used in the neutrino community, namely GENIE, NEUT, NuWro, and GiBUU. In the QE and 2p2h dominant regions, the discrepancy between our measurement and all generator predictions appears as a function of \tmu, whereas in the Res and S/DIS dominant regions, the discrepancy is a function of antimuon angle. 
This underscores the need for improved or additional models in the generators, as well as the challenge of combining the models to form a comprehensive prediction across a broad range of energies.  Our measurement provides strong constraints on these efforts.

This document was prepared by the NOvA Collaboration using the resources of the Fermi National Accelerator Laboratory (Fermilab), a U.S. Department of Energy, Office of Science, HEP User Facility. Fermilab is managed by FermiForward Discovery Group, LLC, acting under Contract No. 89243024CSC000002. This work was supported by the U.S. Department of Energy; the U.S. National Science Foundation; the Department of Science and Technology, India; the European Research Council; the MSMT CR, GA UK, Czech Republic; the RAS, MSHE, and RFBR, Russia; CNPq and FAPEG, Brazil; UKRI, STFC and the Royal Society, United Kingdom; and the state and University of Minnesota. We are grateful for the contributions of the staff of the University of Minnesota at the Ash River Laboratory, and of Fermilab.

\bibliography{NuMuBarCC_Inclusive_Paper_2024}   

\clearpage
 
\onecolumngrid
\section{SUPPLEMENTAL MATERIAL FOR Triple-differential measurement of the muon antineutrino charged-current inclusive cross section using the NOvA Near Detector}

\setcounter{figure}{0}
\setcounter{table}{0}
\setcounter{page}{1}
\setcounter{section}{0}
\setcounter{secnumdepth}{4}
\makeatletter
\renewcommand{\theequation}{S\arabic{equation}}
\renewcommand{\thefigure}{S\arabic{figure}}
\renewcommand{\thetable}{S\arabic{table}}
\renewcommand{\thepage}{S\arabic{page}}
\renewcommand{\thesection}{S-\Roman{section}}

This supplemental material provides additional details supporting the main results presented in the Letter. Section~\ref{sup:purity}
presents the purity and efficiency of the signal in the selected phase-space region. Section~\ref{sup:generators} presents a description of the models used in the neutrino generators used for the comparisons of our results. Finally, Sec.~\ref{sup:chi2n1D} presents comparisons to the generators in the full three-dimensional space of the cross-section measurement, the total cross section for $E_{\nu}$, the single differential cross section in $Q^{2}$, as well as the $\chi^2$ comparisons to all neutrino generators considered in the Letter.

\section{Purity and Efficiency}  
\label{sup:purity}
\begin{figure*}[b]
    \centering
    \includegraphics[width=0.8\linewidth]{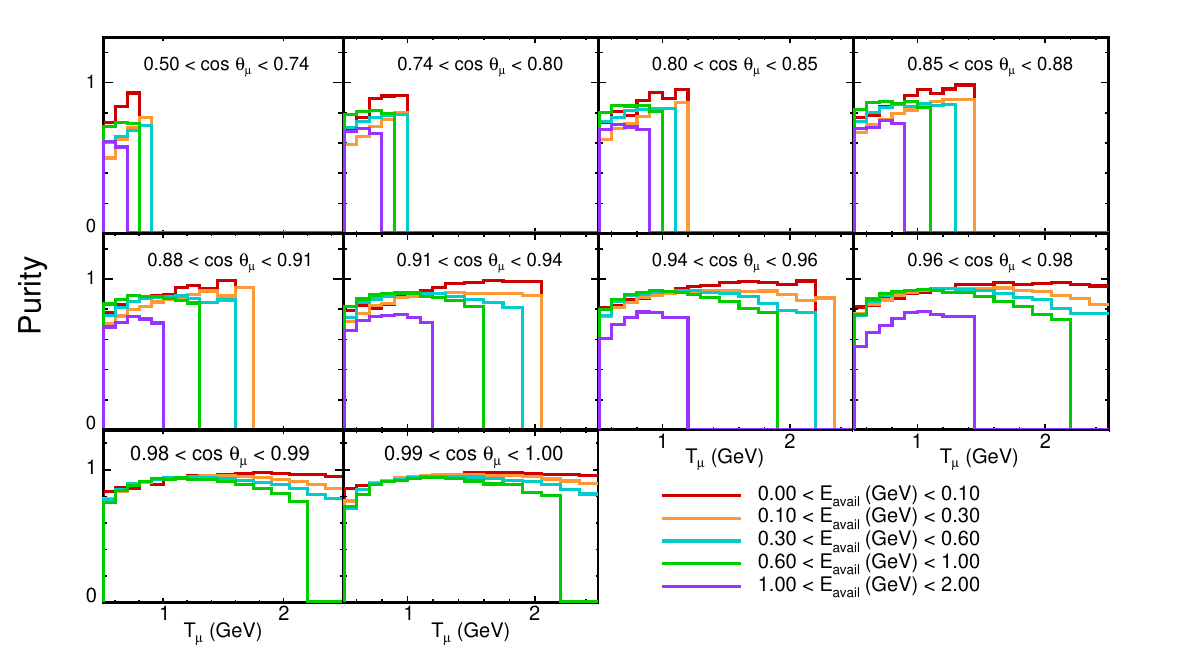}
    \caption{Signal purity vs. of \tmu in panels of $\cos \theta_\mu$. Each color represents a range of \Eavail.}
    \label{fig:purity}
\end{figure*}

\begin{figure*}
    \centering
    \includegraphics[width=0.8\linewidth]{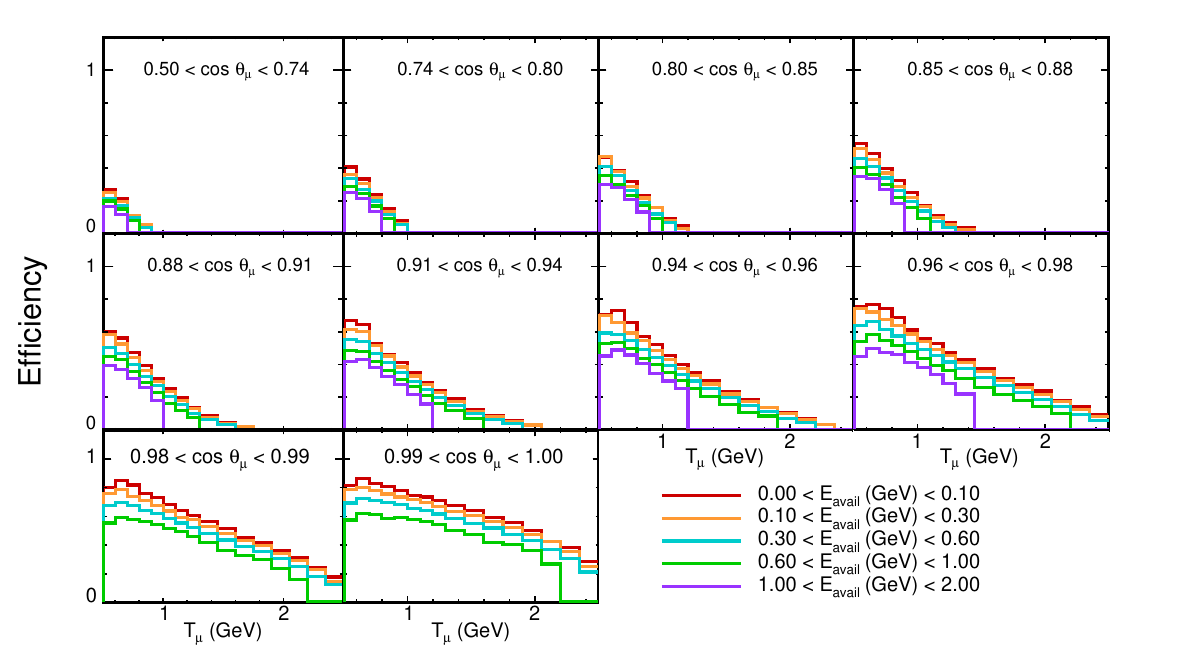}
    \caption{Selection efficiency vs. \tmu in panels of $\cos \theta_\mu$. Each color represents a range of \Eavail.}
    \label{fig:eff}
\end{figure*}

Figures~\ref{fig:purity} and \ref{fig:eff} show the signal purity and selection efficiency in the selected phase-space region. The purity decreases with \Eavail due to an increasing wrong-sign contribution at higher \Eavail, with an additional drop around 1--2\,GeV from NC backgrounds. The sample maintains a high overall purity of about 90\%.

The efficiency decreases with \tmu and at larger scattering angles, primarily due to antimuons escaping containment. At higher \Eavail, increased hadronic activity and showering further reduce the efficiency by complicating antimuon reconstruction. The overall selection efficiency is approximately 32\%.


\section{Brief descriptions of models in generators used for comparison plots} 
\label{sup:generators}
The main body of the paper shows our results compared to a variety of neutrino generators. 
The standard NOvA simulation uses GENIE 3.0.6 \cite{PhysRevD.104.072009, Andreopoulos_2010, Andreopoulos:2015wxa} with the Comprehensive Model Configuration (CMC) \texttt{G18\_10j\_00\_000} and tuned as described in the paper.
We also compare our measurement to neutrino interaction models implemented in GENIE---specifically the tunes \texttt{G18\_10a\_02\_11b}, \texttt{G21\_11a\_00\_000}, and \texttt{AR23\_20i\_00\_000}. 

All three employ the local Fermi gas (LFG) model for the nuclear initial state. The LFG model treats nucleons as non-interacting fermions confined within a potential well, with a momentum distribution up to the Fermi momentum. ``Local" implies that the nuclear density, and thus the Fermi momentum, can vary across the nucleus, accounting for surface effects. 
The spectral-function-enhanced version accounts for correlations between nucleons and the presence of nucleons in excited states or with momenta above the Fermi surface due to short-range correlations.
The \texttt{AR23\_20i\_00\_000} base model does not favor any particular model but provides extensive phase space coverage in missing energy and momentum, making it adaptable for reweighting to many other models, especially including spectral function ones.

For quasielastic (QE) and meson exchange current (2p2h) interactions, {\tt G18\_10a\_02\_11b} uses the Val\`{e}ncia model~\cite{bib:Valencia1,bib:Valencia3,bib:ValenciaGENIE}, {\tt G21\_11a\_00\_000} adopts SuSAv2~\cite{bib:SuSA,bib:SuSAv2}, and {\tt AR23\_20i\_00\_000} uses Val\`{e}ncia for QE and SuSAv2 for 2p2h.
The Val\`{e}ncia model is built from a detailed description of the nuclear many-body system, calculating specific interaction diagrams and incorporating nuclear correlations (such as the random phase approximation, RPA, and 2p2h) explicitly, aiming for a first-principles calculation of the nuclear response. The RPA suppression refers to the reduction in the QE cross section at low momentum transfer due to long-range correlations between nucleons within the nucleus.
The  Val\`{e}ncia model treats QE and 2p2h as separate reaction mechanisms, albeit with strong interconnections and coherent inclusion of nuclear effects. The SuSAv2 approach is rooted in the ``superscaling" phenomenon observed in electron-nucleus scattering, extracting a universal ``scaling function" from electron data that encapsulate nuclear dynamics, and then extends this to neutrino interactions via an assumed factorization of the elementary interaction. This approach naturally blends QE and 2p2h, as the scaling function effectively accounts for both single and multi-nucleon excitations.

Resonant and coherent pion production are modeled using the Berger--Sehgal modification~\cite{PhysRevD.76.113004} to the Rein-Sehgal formalism~\cite{bib:ResReinSeghal}, which relies on external data to tune form factors and phenomenological parameters of the model.
The Rein--Sehgal model does not explicitly include lepton mass effects in the calculation of the differential cross section, with its predictions being significantly higher than recent experimental results. 
The Berger--Sehgal model incorporates lepton mass effects, which gives a significant improvement at lower neutrino energies where the mass of the final-state lepton has a more noticeable impact on the kinematics. The model is tuned on pion-carbon scattering data, so can better account for nuclear effects such as pion absorption, by being fit directly to nuclear data.

Shallow/Deep inelastic scattering (S/DIS) is described by the Bodek--Yang model~\cite{BODEK200270}. 
The model aims to provide a unified description of inelastic scattering from the high-Q$^2$ of the S/DIS regime down to low-Q$^2$ and the resonance region, using effective leading-order parton distribution functions.
The Andreopoulos-Gallagher-Kehayias-Yang (AGKY~\cite{bib:AGKY}) model is used to describe hadronization at low invariant masses (W). At higher W, the hadronization model within PYTHIA6~\cite{bib:PYTHIA} is used. A linear scaling of the two is used for mid-range W.

All CMCs implement the intranuclear cascade (INTRANUKE/hA) model for final-state interactions (FSI).
This simulates the propagation of hadrons through the nuclear medium using the semi-classical stepping method, where the hadrons propagate in a straight line and their reinteraction is determined by the mean free path. The hA model uses one single interaction rather than multiple interactions like the full cascade models; however, this is tuned to external measurement.

Other generators considered include NuWro 21.09.02 (2022) \cite{PhysRevC.86.015505}, NEUT 5.7.0 (2023) \cite{Hayato_2021}, and \texttt{GiBUU} (2023) \cite{BUSS20121,PhysRevD.109.033008}. GiBUU is a transport model that solves a set of coupled semi-classical Boltzmann-Uehling-Uhlenbeck equations. All generators considered use an LFG-based initial nuclear state, with GiBUU applying its own modifications. QE interactions are treated differently: NuWro uses the Llewellyn--Smith formalism~\cite{bib:LlewellynSmith}, which uses a Relativistic Fermi Gas approach to simulate the initial state, and tends to overpredict the cross section at very low momentum due to the lack of RPA suppression. NEUT uses Val\`{e}ncia, and GiBUU incorporates RPA corrections with dipole form factors. 
2p2h modeling in NuWro and NEUT relies on Val\`{e}ncia, while GiBUU uses a phenomenological model based on semi-inclusive electron-scattering data and dipole form factors.

For resonance production, NuWro applies a custom model, NEUT uses the Berger--Sehgal modifications to the Rein--Sehgal formalism, and GiBUU uses the MAID model including electromagnetic form factors. S/DIS is modeled via Bodek--Yang in NuWro and NEUT, and through a data-driven approach in GiBUU. FSI treatments vary: NuWro uses its own cascade model, NEUT applies a semi-classical intranuclear cascade, and GiBUU employs a full BUU transport framework.
GiBUU models all relevant reaction channels (QE, Res, S/DIS, 2p2h) and handles the FSI within the same consistent transport framework. The particles produced in the initial interaction are immediately propagated through the nuclear potential, undergoing further collisions and interactions until they exit the nucleus. 

Table~\ref{tab:generators} shows a summary of various models used in the generators used in the analysis.

\begin{table}[htbp]
\centering
\caption{Neutrino interaction modeling components in GENIE 3.0.6 \cite{PhysRevD.104.072009, Andreopoulos_2010, Andreopoulos:2015wxa} CMCs and in
\texttt{NuWro~21.09.02} (2022)~\cite{PhysRevC.86.015505}, NEUT 5.7.0 (2023) \cite{Hayato_2021}, and \texttt{GiBUU} (2023) \cite{BUSS20121,PhysRevD.109.033008}.
}
\setlength\extrarowheight{2pt}
\begin{tabular}{lcccccc}
\hline\hline
\textbf{Generator}          & \textbf{Initial State} & \textbf{QE}       & \textbf{2p2h}       & \textbf{Resonance} & \textbf{S/DIS}               & \textbf{FSI}       \\
\hline
\multirow{ 2}{*}{G18\_10a\_02\_11b}  & \multirow{ 2}{*}{LFG}  & \multirow{ 2}{*}{Val\`{e}ncia}  & \multirow{ 2}{*}{Val\`{e}ncia} & \multirow{ 2}{*}{Berger--Sehgal} & Bodek--Yang & \multirow{ 2}{*}{INTRANUKE/hA} \\
 &                   &         &          &     & + PYTHIA &       \\
\multirow{ 2}{*}{G21\_11a\_00\_000} & \multirow{ 2}{*}{LFG} & \multirow{ 2}{*}{SuSAv2} & \multirow{ 2}{*}{SuSAv2}  & \multirow{ 2}{*}{Berger--Sehgal}  & Bodek--Yang & \multirow{ 2}{*}{INTRANUKE/hA}       \\
          &                   &            &            &     & + PYTHIA  &      \\

\multirow{ 2}{*}{Ar23\_20i\_00\_000} & Spectral & \multirow{ 2}{*}{Val\`{e}ncia} & \multirow{ 2}{*}{SuSAv2} & \multirow{ 2}{*}{Berger--Sehgal}     & Bodek--Yang & \multirow{ 2}{*}{INTRANUKE/hA} \\ 
          &    LFG               &            &            &     & + PYTHIA  &      \\
\hline
\multirow{ 2}{*}{NuWro 21.09.02} & \multirow{ 2}{*}{LFG} & \multirow{ 2}{*}{Llewellyn--Smith} & \multirow{ 2}{*}{Val\`{e}ncia} & \multirow{ 2}{*}{Custom NuWro} & \multirow{ 2}{*}{Bodek--Yang}       & \multirow{ 2}{*}{NuWro cascade}   \\
          &                   &            &            &     &   &      \\
\multirow{ 2}{*}{NEUT 5.7.0}  & \multirow{ 2}{*}{LFG} & \multirow{ 2}{*}{Val\`{e}ncia} & \multirow{ 2}{*}{Val\`{e}ncia} & \multirow{ 2}{*}{Berger-Sehgal}  & \multirow{ 2}{*}{Bodek--Yang} & INC    \\
  &   &   &   &   &   &   (semi-classical)   \\
\multirow{ 2}{*}{GiBUU (2023) } & Modified  & Dipole FF  & Electron   & MAID  & \multirow{ 2}{*}{Data-driven} & \multirow{ 2}{*}{BUU transport}  \\
 & LFG  & + RPA  &  scattering data   & + EM FF       &       &           \\
\hline\hline
\end{tabular}
\label{tab:generators}
\end{table}

\section{3D comparisons, 1D results and $\chi^2$ comparisons} 
\label{sup:chi2n1D}

\begin{figure*}
    \centering
    \includegraphics[width=1.0\linewidth]{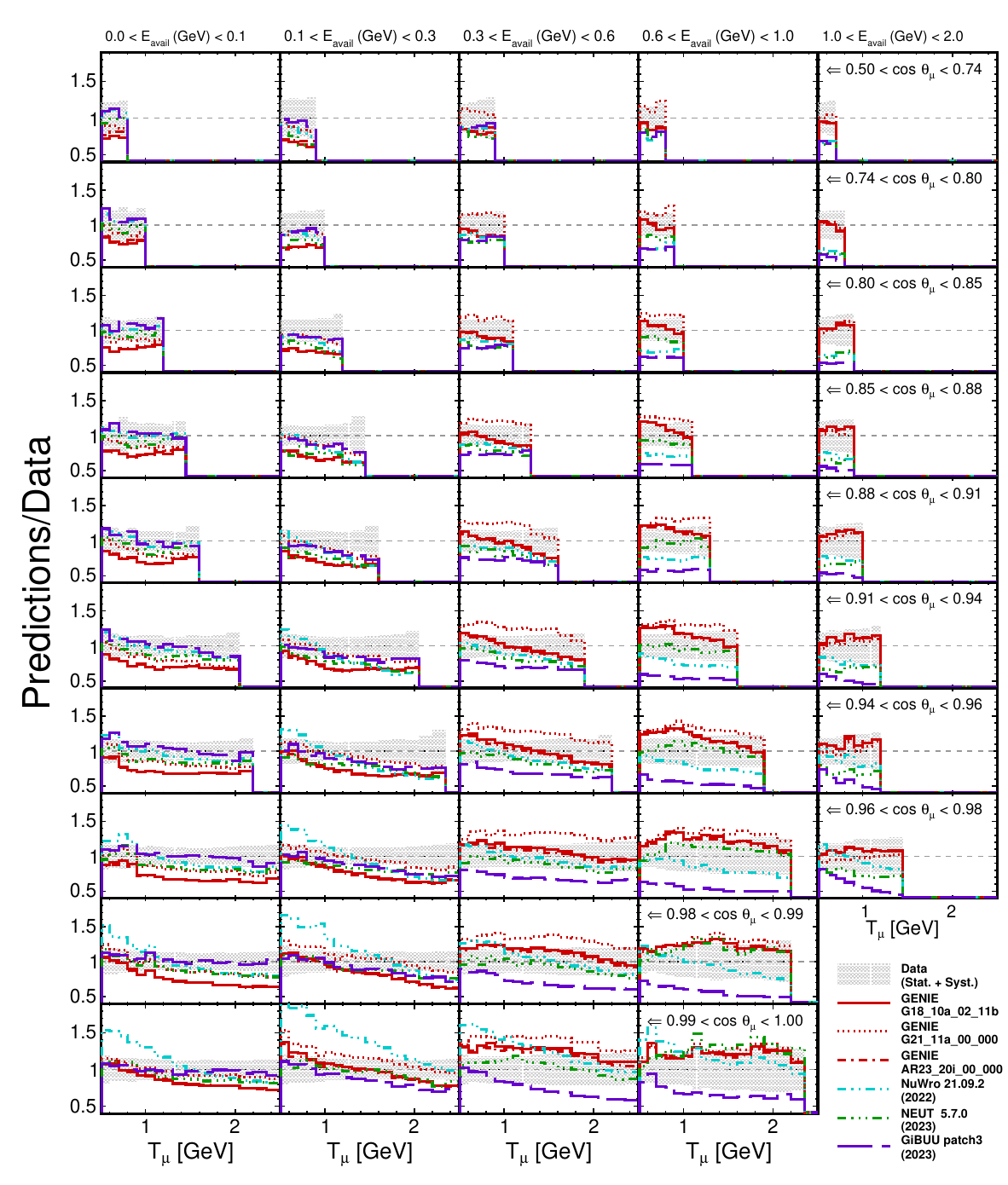}
    \caption{Comparisons of our measurements to various GENIE CMCs, and the neutrino generators NuWro 21.09.02 (2022) \cite{PhysRevC.86.015505}, NEUT 5.7.0 (2023) \cite{Hayato_2021}, and GiBUU (2023) \cite{BUSS20121, PhysRevD.109.033008}. See Tab.~\ref{tab:generators} for a summary of the models used in these generators. Each plot shows the ratio of the prediction to our measurement vs. \tmu, with each column representing one of the \Eavail bins, and each row representing one of the \cosmu bins.  }
    \label{fig:comparison}
\end{figure*}

Figure~\ref{fig:comparison} shows comparisons of our measurements to generators mentioned above. Each plot shows the ratio of the prediction to our measurement vs. \tmu, with each column representing one of the \Eavail bins, and each row representing one of the \cosmu bins.  

\begin{figure}
    \centering
    \includegraphics[width=0.4\linewidth]{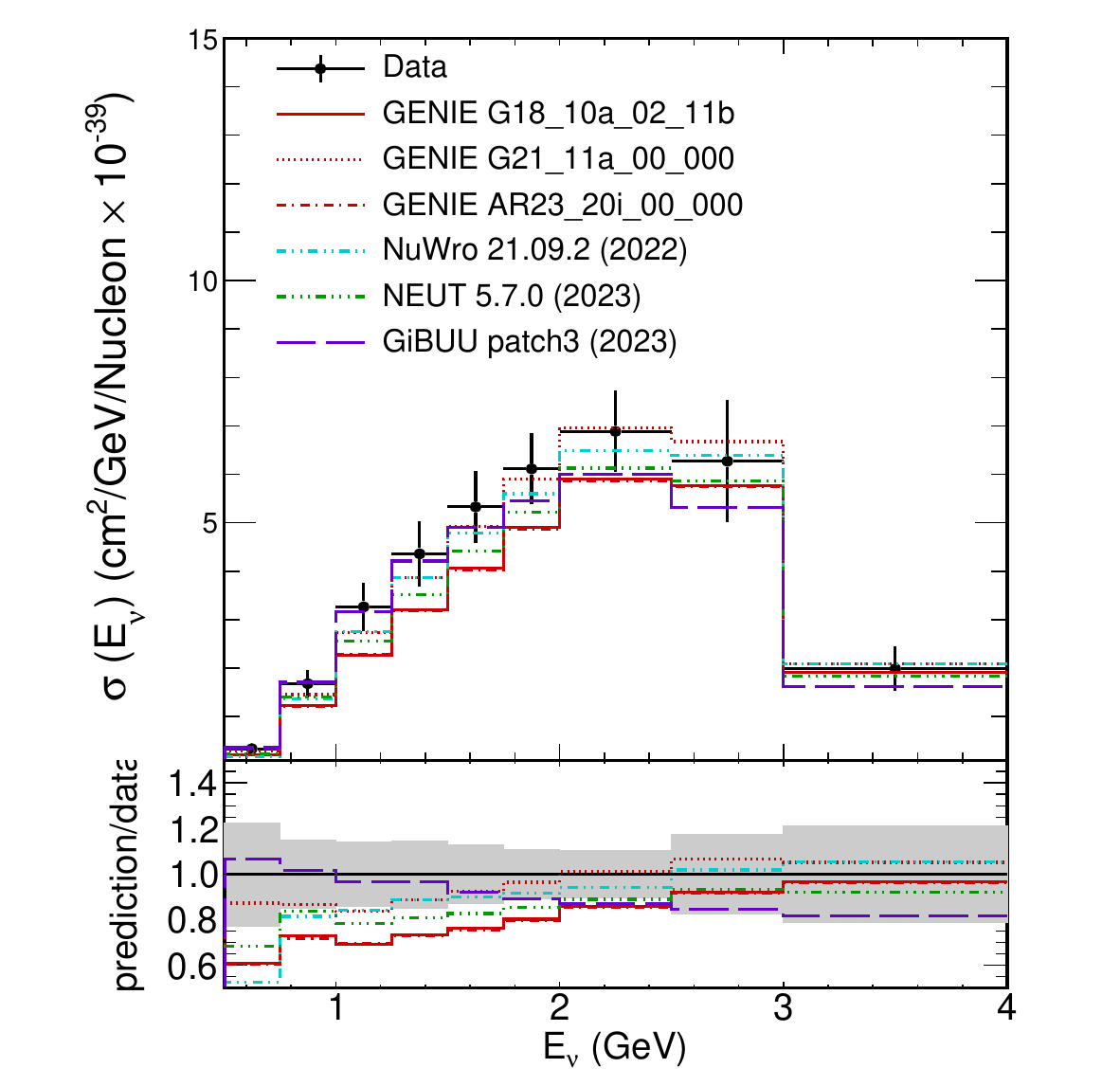} \,
   \includegraphics[width=0.4\linewidth]{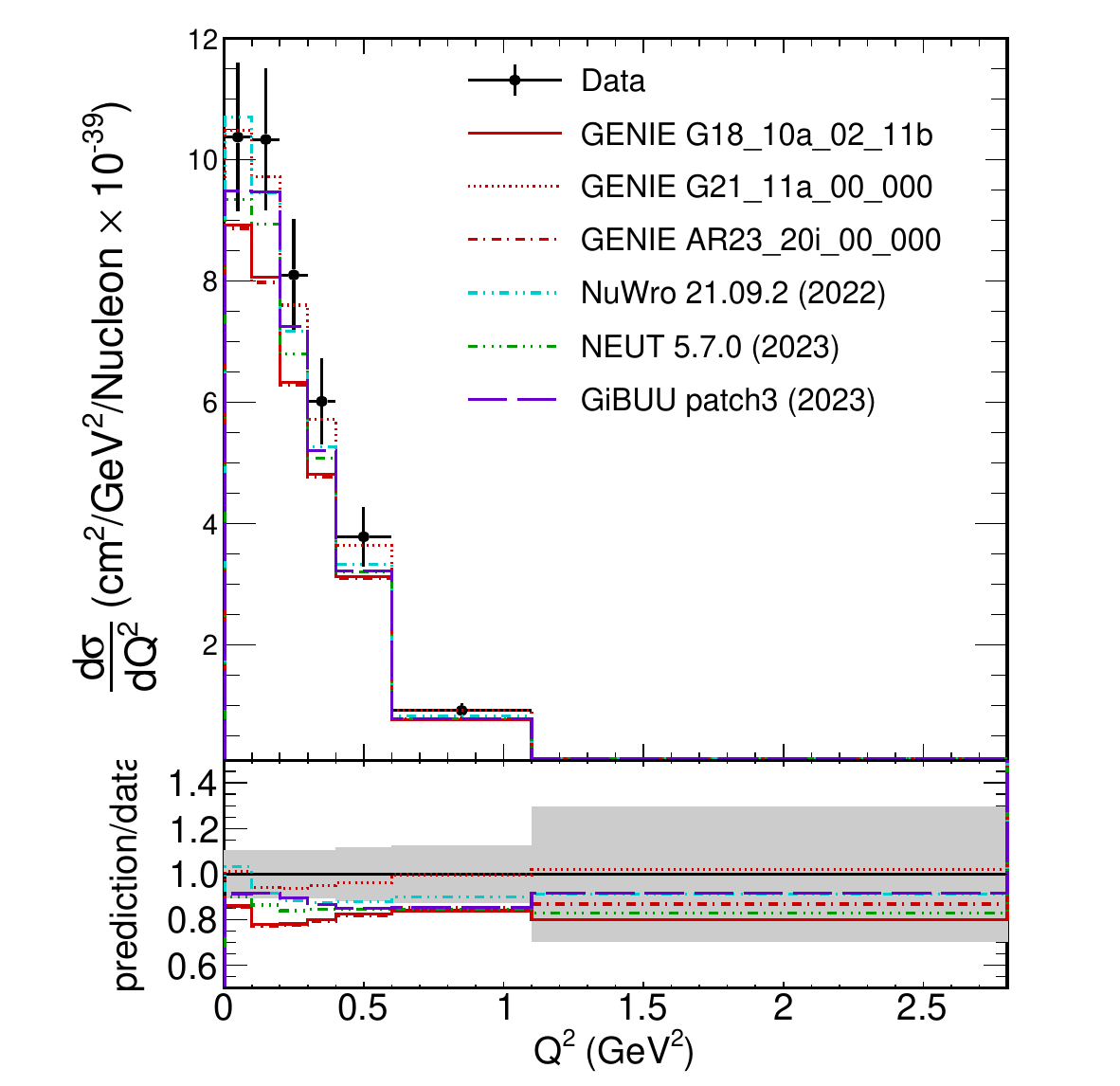}
 
   \caption{The neutrino cross section vs. $E_{\nu}$ (left) and the single-differential cross section vs. four-momentum transfer $Q^{2}$. Both cross sections are extracted in the phase space defined by the antimuon kinematics in the main analysis.}
    \label{fig:results_1D}
\end{figure}

The total cross section for $E_{\nu}$ is shown in Fig.~\ref{fig:results_1D}(left), and a single differential cross section in $Q^{2}$ is shown in Fig.~\ref{fig:results_1D}(right). 
These are calculated only in the phase space defined by our antimuon kinematics measurement, which causes $\sigma(E_\nu)$ to deviate from the simple linear growth with \enu. For the $E_{\nu}$ comparisons, at low $E_{\nu}$, all generators are under-predicting data except GiBUU. For all generators, the shape agreement in energy is good and the disagreements with data are mostly in the normalization. All generator predictions are under-predicting data for the $Q^{2}$ comparisons. 

Table \ref{table:chisq} summarizes the $\chi^{2}$/ndof values for each model compared to data. The $\chi^{2}$ values are calculated using covariance matrices to account for bin-to-bin correlations. Overall, the $\chi^{2}$ values are strongly driven by the \Eavail region below 0.1\,GeV, which has about half of our selected events. 

\begin{table}[]
\centering
\caption{$\chi^{2}$ calculated for each model compared to data cross sections. Degrees of freedom are 459, 9, and 7, for triple-differential, $E_{\nu}$, and $Q^{2}$ results, respectively. The shape only $\chi^{2}$ values are shown in parenthesis.}
\label{table:chisq}
\setlength\extrarowheight{4pt}
\begin{tabular}{lccc}
\hline
Generator             & $d\sigma^3/ dT_\mu d\cos\theta_{\mu} dE$ & $\sigma(E_{\nu}) / E_{\nu} $ & 
$ d\sigma / dQ^{2} $ \\ \hline
GENIE v3.0.6                & 9.5 (8.4)   & 0.9 (0.8)  & 9.2 (9.0)    \\
GENIE v3.0.6 NOvA-tune-v2   & 2.9 (2.9)   & 0.3 (0.5)  & 1.3 (1.4)    \\
GENIE v3.4.0                & 11.5 (8.3)  & 1.0 (1.1)  & 13.9 (10.9)  \\
NEUT 5.7.0 (2023)           & 9.3 (7.5)   & 0.8 (0.9)  & 2.8 (2.5)    \\
NuWRO 21.09.02 (2022)       & 15.3 (13.8) & 0.8 (0.7)  & 15.4 (14.0)  \\
GiBUU (2023)                & 4.8 (4.3)   & 0.9 (1.0)  & 1.0 (0.8)    \\ \hline
\end{tabular}%
\end{table}


\end{document}

%% file: novatriplecross2024.tex
\newcommand{\ANL}{Argonne National Laboratory, Argonne, Illinois 60439, 
USA}
\newcommand{\Bandirma}{Bandirma Onyedi Eyl\"ul University, Faculty of 
Engineering and Natural Sciences, Engineering Sciences Department, 
10200, Bandirma, Balıkesir, Turkey}
\newcommand{\ICS}{Institute of Computer Science, The Czech 
Academy of Sciences, 
182 07 Prague, Czech Republic}
\newcommand{\IOP}{Institute of Physics, The Czech 
Academy of Sciences, 
182 21 Prague, Czech Republic}
\newcommand{\Atlantico}{Universidad del Atlantico,
Carrera 30 No.\ 8-49, Puerto Colombia, Atlantico, Colombia}
\newcommand{\BHU}{Department of Physics, Institute of Science, Banaras 
Hindu University, Varanasi, 221 005, India}
\newcommand{\UCLA}{Physics and Astronomy Department, UCLA, Box 951547, Los 
Angeles, California 90095-1547, USA}
\newcommand{\Caltech}{California Institute of 
Technology, Pasadena, California 91125, USA}
\newcommand{\Cochin}{Department of Physics, Cochin University
of Science and Technology, Kochi 682 022, India}
\newcommand{\Charles}
{Charles University, Faculty of Mathematics and Physics,
 Institute of Particle and Nuclear Physics, Prague, Czech Republic}
\newcommand{\Cincinnati}{Department of Physics, University of Cincinnati, 
Cincinnati, Ohio 45221, USA}
\newcommand{\CSU}{Department of Physics, Colorado 
State University, Fort Collins, CO 80523-1875, USA}
\newcommand{\CTU}{Czech Technical University in Prague,
Brehova 7, 115 19 Prague 1, Czech Republic}
\newcommand{\Dallas}{Physics Department, University of Texas at Dallas,
800 W. Campbell Rd. Richardson, Texas 75083-0688, USA}
\newcommand{\DallasU}{University of Dallas, 1845 E 
Northgate Drive, Irving, Texas 75062 USA}
\newcommand{\Delhi}{Department of Physics and Astrophysics, University of 
Delhi, Delhi 110007, India}
\newcommand{\JINR}{Joint Institute for Nuclear Research,  
Dubna, Moscow region 141980, Russia}
\newcommand{\Erciyes}{
Department of Physics, Erciyes University, Kayseri 38030, Turkey}
\newcommand{\FNAL}{Fermi National Accelerator Laboratory, Batavia, 
Illinois 60510, USA}
\newcommand{\FSU}{Florida State University, Tallahassee, Florida 32306, USA}
\newcommand{\UFG}{Instituto de F\'{i}sica, Universidade Federal de 
Goi\'{a}s, Goi\^{a}nia, Goi\'{a}s, 74690-900, Brazil}
\newcommand{\Guwahati}{Department of Physics, IIT Guwahati, Guwahati, 781 
039, India}
\newcommand{\Harvard}{Department of Physics, Harvard University, 
Cambridge, Massachusetts 02138, USA}
\newcommand{\Houston}{Department of Physics, 
University of Houston, Houston, Texas 77204, USA}
\newcommand{\IHyderabad}{Department of Physics, IIT Hyderabad, Hyderabad, 
502 205, India}
\newcommand{\Hyderabad}{School of Physics, University of Hyderabad, 
Hyderabad, 500 046, India}
\newcommand{\IIT}{Illinois Institute of Technology,
Chicago IL 60616, USA}
\newcommand{\Imperial}{Imperial College London, Department of Physics,
 London, United Kingdom}
\newcommand{\Indiana}{Indiana University, Bloomington, Indiana 47405, 
USA}
\newcommand{\INR}{Institute for Nuclear Research of Russia, Academy of 
Sciences 7a, 60th October Anniversary prospect, Moscow 117312, Russia}
\newcommand{\UIowa}{Department of Physics and Astronomy, University of Iowa, 
Iowa City, Iowa 52242, USA}
\newcommand{\ISU}{Department of Physics and Astronomy, Iowa State 
University, Ames, Iowa 50011, USA}
\newcommand{\Irvine}{Department of Physics and Astronomy, 
University of California at Irvine, Irvine, California 92697, USA}
\newcommand{\Jammu}{Department of Physics and Electronics, University of 
Jammu, Jammu Tawi, 180 006, Jammu and Kashmir, India}
\newcommand{\Lebedev}{Nuclear Physics and Astrophysics Division, Lebedev 
Physical 
Institute, Leninsky Prospect 53, 119991 Moscow, Russia}
\newcommand{\Magdalena}{Universidad del Magdalena, Carrera 32 No 22-08 Santa Marta, Colombia}
\newcommand{\MSU}{Department of Physics and Astronomy, Michigan State 
University, East Lansing, Michigan 48824, USA}
\newcommand{\Crookston}{Math, Science and Technology Department, University 
of Minnesota Crookston, Crookston, Minnesota 56716, USA}
\newcommand{\Duluth}{Department of Physics and Astronomy, 
University of Minnesota Duluth, Duluth, Minnesota 55812, USA}
\newcommand{\Minnesota}{School of Physics and Astronomy, University of 
Minnesota Twin Cities, Minneapolis, Minnesota 55455, USA}
\newcommand{\Mississippi}{University of Mississippi, University, Mississippi 38677, USA}
\newcommand{\NISER}{National Institute of Science Education and Research, An OCC of Homi Bhabha
National Institute, Bhubaneswar, Odisha, India}
\newcommand{\OSU}{Department of Physics, Ohio State University, Columbus,
Ohio 43210, USA}
\newcommand{\Oxford}{Subdepartment of Particle Physics, 
University of Oxford, Oxford OX1 3RH, United Kingdom}
\newcommand{\Panjab}{Department of Physics, Panjab University, 
Chandigarh, 160 014, India}
\newcommand{\Pitt}{Department of Physics, 
University of Pittsburgh, Pittsburgh, Pennsylvania 15260, USA}
\newcommand{\QMU}{Particle Physics Research Centre, 
Department of Physics and Astronomy,
Queen Mary University of London,
London E1 4NS, United Kingdom}
\newcommand{\RAL}{Rutherford Appleton Laboratory, Science 
and 
Technology Facilities Council, Didcot, OX11 0QX, United Kingdom}
\newcommand{\SAlabama}{Department of Physics, University of 
South Alabama, Mobile, Alabama 36688, USA} 
\newcommand{\Carolina}{Department of Physics and Astronomy, University of 
South Carolina, Columbia, South Carolina 29208, USA}
\newcommand{\SDakota}{South Dakota School of Mines and Technology, Rapid 
City, South Dakota 57701, USA}
\newcommand{\SMU}{Department of Physics, Southern Methodist University, 
Dallas, Texas 75275, USA}
\newcommand{\Stanford}{Department of Physics, Stanford University, 
Stanford, California 94305, USA}
\newcommand{\Sussex}{Department of Physics and Astronomy, University of 
Sussex, Falmer, Brighton BN1 9QH, United Kingdom}
\newcommand{\Syracuse}{Department of Physics, Syracuse University,
Syracuse NY 13210, USA}
\newcommand{\Tennessee}{Department of Physics and Astronomy, 
University of Tennessee, Knoxville, Tennessee 37996, USA}
\newcommand{\Texas}{Department of Physics, University of Texas at Austin, 
Austin, Texas 78712, USA}
\newcommand{\Tufts}{Department of Physics and Astronomy, Tufts University, Medford, 
Massachusetts 02155, USA}
\newcommand{\UCL}{Physics and Astronomy Department, University College 
London, 
Gower Street, London WC1E 6BT, United Kingdom}
\newcommand{\Virginia}{Department of Physics, University of Virginia, 
Charlottesville, Virginia 22904, USA}
\newcommand{\WSU}{Department of Mathematics, Statistics, and Physics,
 Wichita State University, 
Wichita, Kansas 67260, USA}
\newcommand{\WandM}{Department of Physics, William \& Mary, 
Williamsburg, Virginia 23187, USA}
\newcommand{\Wisconsin}{Department of Physics, University of 
Wisconsin-Madison, Madison, Wisconsin 53706, USA}
\newcommand{\deceased}{Deceased.}
\affiliation{\ANL}
\affiliation{\Atlantico}
\affiliation{\Bandirma}
\affiliation{\BHU}
\affiliation{\Caltech}
\affiliation{\Charles}
\affiliation{\Cincinnati}
\affiliation{\Cochin}
\affiliation{\CSU}
\affiliation{\CTU}
\affiliation{\Delhi}
\affiliation{\Erciyes}
\affiliation{\FNAL}
\affiliation{\FSU}
\affiliation{\UFG}
\affiliation{\Guwahati}
\affiliation{\Houston}
\affiliation{\Hyderabad}
\affiliation{\IHyderabad}
\affiliation{\IIT}
\affiliation{\Imperial}
\affiliation{\Indiana}
\affiliation{\ICS}
\affiliation{\INR}
\affiliation{\IOP}
\affiliation{\UIowa}
\affiliation{\ISU}
\affiliation{\Irvine}
\affiliation{\JINR}
\affiliation{\Magdalena}
\affiliation{\MSU}
\affiliation{\Duluth}
\affiliation{\Minnesota}
\affiliation{\Mississippi}
\affiliation{\OSU}
\affiliation{\NISER}
\affiliation{\Panjab}
\affiliation{\Pitt}
\affiliation{\QMU}
\affiliation{\SAlabama}
\affiliation{\Carolina}
\affiliation{\SMU}
\affiliation{\Sussex}
\affiliation{\Syracuse}
\affiliation{\Texas}
\affiliation{\Tufts}
\affiliation{\UCL}
\affiliation{\Virginia}
\affiliation{\WSU}
\affiliation{\WandM}
\affiliation{\Wisconsin}

\author{S.~Abubakar}
\affiliation{\Erciyes}

\author{M.~A.~Acero}
\affiliation{\Atlantico}

\author{B.~Acharya}
\affiliation{\Mississippi}

\author{P.~Adamson}
\affiliation{\FNAL}









\author{N.~Anfimov}
\affiliation{\JINR}


\author{A.~Antoshkin}
\affiliation{\JINR}


\author{E.~Arrieta-Diaz}
\affiliation{\Magdalena}

\author{L.~Asquith}
\affiliation{\Sussex}


\author{A.~Aurisano}
\affiliation{\Cincinnati}






\author{N.~Balashov}
\affiliation{\JINR}

% ...

\collaboration{The NOvA Collaboration}
\noaffiliation


